\documentclass[12pt]{JHEP3}
\usepackage{graphicx}
\usepackage{epsfig}
\usepackage{amssymb}
\usepackage{bm}
\usepackage{amsmath}
\newcommand{\bea}{\begin{eqnarray}}
\newcommand{\eea}{\end{eqnarray}}

\makeatletter
\@addtoreset{equation}{section}
\makeatother

\title{
Possibility of ferromagnetic neutron matter
}
\author{
Koji Hashimoto
\\

{\it Department of Physics, Osaka University,
Toyonaka, Osaka 560-0043, Japan}\\
{\it Mathematical Physics Lab., RIKEN Nishina Center,
Saitama 351-0198, Japan}\\
E-mail: \email{koji(at)phys.sci.osaka-u.ac.jp}\\ 
}

\abstract{
We study ferromagnetism at high density of neutrons in the 
QCD hadron phase, by using the simplest chiral effective model incorporating
magnetic fields and the chiral anomaly. Under the assumption of spatial 
homogeneity, we calculate the energy density as a function of neutron density,
with a magnetization and a neutral pion condensation {\it a la} Dautry and Neyman. 
We find that at a high density
the energy of the ferromagnetic order 
is lower than that of the ordinary neutron matter, and the reduction 
effect is enhanced by the anomaly. Compared to the inhomogeneous 
phase with the alternating layer structure, our ferromagnetic phase turns out to be
unfavored. However, once an axial vector meson condensation 
is taken into account in our simplest model, 
the ferromagnetic energy density is lowered significantly,
which still leaves some room for a possible realization of a QCD ferromagnetic phase
and ferromagnetic magnetars.
}

\preprint{
{\normalsize OU-HET-844} \\
{\normalsize RIKEN-MP-104}
}


\begin{document}
\setcounter{page}{1}

\section{Motivation: QCD and ferromagnetism}

Ferromagnetic order in nature always attracts interest for study as it manifests microscopic structure of matter and materials. 
Among observed magnetic fields in nature, perhaps the strongest stable magnetic field is on the surface of magnetars, which goes up to $10^{15}$ [G] and more \cite{EnotoProc,Harding:2006qn,Mereghetti:2008}. The mechanism for generating such a strong field is 
yet to be uncovered,
and it is natural to resort the origin to the high density of neutrons of which the neutron stars consist. In fact, 
after the discovery of pulsars, the possibility of ferromagnetism at neutron stars was proposed \cite{Brownell:1969,Rice,Silverstein,Maxima}. 
However, 
numerical simulations of neutron matter with realistic inter-nucleon potentials have
not shown the ferromagnetic phase \cite{Bordbar:2008}. So the possibility of the ferromagnetic phase at high density neutron matter, if exists in nature, 
waits for a new mechanism of the spontaneous magnetization.

In this paper, we study the possibility of the ferromagnetic phase at high density of neutrons, by using the simplest but general chiral effective action. Low energy dynamics of neutrons is governed by the chirally symmetric interactions through pions
and the spin-magnetic coupling with magnetic fields.
Our model consists of dense neutrons coupled with neutral pions and magnetic fields,
together with the chiral anomaly term. These are indispensable ingredients,
and we will see the outcome for the magnetic phase from this minimal model.

The reason for choosing the neutral pion is simply for the realization of the 
ferromagnetism, as
other pion condensations such as charged pion condensation 
\cite{Baym:1975tm,Baym:1978sz} have not been shown to
exhibit a ferromagnetism.
In addition,
with a neutral pion condensation of the form
$\Pi_0(x) \propto \sin \mbox{\boldmath $k$} \cdot \mbox{\boldmath $x$}$,
a neutron lattice is formed with 
an alternating layer structure (ALS) \cite{Tamagaki:1976ds,Takatsuka:1977uj,Takatsuka:1978ku,Takatsuka:1993pv},
then the neutron spins cancel each other, and macroscopic magnetization would not emerge.
In this paper, instead, we analyze a neutral pion condensation of the different form
$\Pi_0(x) =\mbox{\boldmath $q$} \cdot \mbox{\boldmath $x$}$ following
Dautry and Neyman \cite{DN}\footnote{For a recent review on the condensation (called chiral density wave), see \cite{Buballa:2014tba}.}, and generalize the study to include magnetic fields
and QCD anomaly.

Our study is motivated by the earlier work
\cite{Eto:2012qd} in which, together with M.~Eto and T.~Hatsuda, 
the author proposed 
a mechanism for a ferromagnetic phase at high density of neutrons.
The mechanism utilizes  a neutral pion domain wall \cite{Hatsuda:1986nq} coupled
to the magnetic field through the QCD chiral anomaly \cite{Son:2007ny}.
A spontaneous magnetization was shown in \cite{Eto:2012qd} in the approximation
of a single wall and  one-loop neutrons.
In this paper, we generalize the idea, and study in the simplest chiral model 
the
Fermi energy of the dense neutrons and its back-reaction due to the pion condensation and the magnetic fields.
A successive array of the domain walls can be approximated by the
linear pion condensation of Dautry and Neyman. 

Let us describe what we will find in this paper.
\begin{itemize}
\item Toy model of neutral fermions. \\[3pt]
First we provide
a toy model of a neutral fermion with a Zeeman coupling to magnetic fields. Under the assumption of the spatial homogeneity, we calculate the
energy density of the ferromagnetic phase and show that it is favored compared to 
the ordinary fermion matter. (Sec.~2)
\item
Simplest chiral model and ferromagnetic order. \\[3pt]
The toy model 
of the neutral fermions 
is the essential part
of the chiral model of neutrons and pions. We 
analyze the simplest chiral effective model of dense neutrons and neutral pions,
together with the magnetic field coupling and the QCD anomaly. We find that 
the neutral pion condensation
of form proposed by Dautry and Neyman is precisely in the same place as the magnetization,
under the assumption of the spatial homogeneity. The energy density of the 
ferromagnetic-pion-condensation phase is lower than the ordinary neutron matter 
at high density around $\rho > 5\rho_0$ where $\rho_0$ is 
the standard nuclear density. Furthermore, 
the chiral anomaly term actually helps the ferromagnetic order.
The generated magnetic field is $\sim 40$ [MeV] $\sim {\cal O}(10^{17})$[G].
(Sec.~3.1, 3.2)
\item Comparison to ALS.\\[3pt]
We compare our energy density with that of the inhomogeneous ALS phase
(which does not exhibit a magnetization),
and find that the ALS phase is favored. The energy gain of the ALS is by several times 
greater than that of our ferromagnetic phase. (Sec.~3.3)
\item Axial vector meson condensation.\\[3pt]
To seek for the possibility of the ferromagnetism, we look at the {\it axial vector meson condensation} accompanied by our model. Indeed, any axial vector meson 
plays the same role as the neutral pions, and the axial vector meson
condensation further reduces the
energy density of the ferromagnetic phase significantly. Incorporation of a higher
vector meson tower and its condensation is studied by using the AdS/CFT correspondence. (Sec.~3.4)
\end{itemize}
In summary, we analyze the ferromagnetic order of our simplest chiral model of dense neutrons with magnetic fields and the QCD anomaly. We find that our ferromagnetic order, as its simplest form, is not favored compared to the ALS phase. 
We further find that the axial vector meson condensation and the QCD anomaly,
together with the pion condensation of Dautry and Neyman, significantly
helps the reduction of the energy density, which suggests a necessity for
further investigation for a
realization of the ferromagnetic phase.
The analysis in this paper is for the minimal model as we have emphasized above, 
so the result
should be understood only qualitatively. Incorporation of realistic nuclear forces
and nucleon contact terms, and also inclusion of electrons and protons, would be
important for a further progress for realizing the QCD ferromagnetic phase
at high density of neutrons.

The organization of this paper is as follows. In Sec.~2, we provide the toy model
of neutral fermions and study a ferromagnetism at high density. In Sec.~3.1 and Sec.~3.2,
we analyze the simplest chiral model of neutrons with the pions, magnetic fields
and the chiral anomaly. In Sec.~3.3, we present our result on the energy plot and a  
comparison to the ALS phase is made. In Sec.~3.4 and 3.5, incorporation of
the axial vector meson condensation is studied, with a help of the AdS/CFT correspondence.  Sec.~4 is for a summary and discussions.


\section{Toy model of dense neutral fermions}

\subsection{Fermions, magnetization and constant magnetic field}

We are interested in the effect of spin and its magnetization, for a general fermion system. 
The magnetization is a condensation of a spin operator of fermions.
Since for relativistic systems the spin operator of a fermion $\psi(x)$ 
is given by a spatial component of an axial current,
\begin{eqnarray}
{\cal S}_i(x) = \frac12 \bar{\psi} \gamma_i \gamma_5 \psi \, ,
\end{eqnarray}
we can systematically write an action for the fermion with the spin magnetic coupling. 
We consider 
the following general system of a neutral fermion. It is 
a system of a free neutral
fermion $\psi(x)$ with a mass $m$ in 4 spacetime dimensions, 
with a Zeeman coupling under a dynamical magnetic field $B_i$,
\begin{eqnarray}
{\cal L}_{\rm fermion} &= &\bar{\psi} \left(
i \partial_\mu \gamma^\mu - m + i\gamma_0 \mu 
\right)\psi 
+\alpha \; \bar{\psi} \gamma_i \gamma_5 \psi \; B_i
-\frac12 B_i^2 \, .
\label{lagful}
\end{eqnarray}
A chemical potential $\mu$ for the fermion number is introduced such that we can
treat the fermion density $\rho$.
Since the magnetization is a back-reaction to the spacetime magnetic field, we have 
included the kinetic term for the magnetic field $B_i$.
The second term in (\ref{lagful}) is the Zeeman coupling $\alpha$
between the spin of the fermion and the magnetic field $B_i$. 
The Zeeman coupling is a part of a so-called 
Pauli term.\footnote{
One would notice that the interaction term added in (\ref{lagful}) does not respect 
the Lorentz invariance,
as only the spatial index $i$ is summed. However, since we need the density for our analysis, the
chemical potential term already broke the Lorentz invariance in (\ref{lagful}), 
so we need not worry about it.}

Our fermion does not have a charge, since we are interested in effects induced particularly by 
the spin magnetic interaction.\footnote{For charged fermions, the magnetic field provides Landau levels which may change the story quite a bit, and will bring an interesting outcome.
We shall come back to the charged fermion case elsewhere.}
So in our model 
there is no standard canonical coupling between the gauge field for the magnetic field and the fermion $\psi$. Normally,
for a charged spin-$1/2$ fermion 
with an electric charge $e$,
 the Zeeman coupling
is measured in the unit of a Bohr magneton, as 
\begin{eqnarray}
\alpha = \frac{g}{2} \frac{e}{2m}
\label{alpha}
\end{eqnarray}
where $g$ is the ``$g$-factor" and $e/2m$ is the Bohr magneton.  
Our fermion does not have the electric charge, so we shall treat $\alpha$ as a general
spin-magnetic coupling. The relation
(\ref{alpha}) can be thought of as a reference, for example for a neutron which will be treated in section 3.

Under a constant magnetic field $B_i$, we consider the behavior of the dense 
neutral fermions.
We shall quantize the spin of the fermion along the direction of the magnetic field. Then
there are two Fermi seas, one is for the up spin and the other is for the down spin.
In the presence of the background magnetic field $B_i$, due to the spin-magnetic 
Zeeman coupling, we have a Zeeman splitting
for the Fermi energy for spin up and down states.

It is easy to evaluate the free energy of each spin sector. In the non-relativistic fermions
where mass $m$ is large compared to the Fermi energy of the fermions, we 
obtain
\begin{eqnarray}
F_{\uparrow} = -\frac{(2 m)^{3/2}}{15 \pi^2}
\left(
\mu - m - \alpha B
\right)^{5/2} \, \quad
F_{\downarrow} = -\frac{(2 m)^{3/2}}{15 \pi^2}
\left(
\mu - m + \alpha B
\right)^{5/2}\, .
\label{freefer}
\end{eqnarray}
The difference is just the sign of the Zeeman coupling, due to the spins.
We have denoted $B$ as the magnitude of the magnetic field $B_i$.
The total free energy of the system, including the magnetic field energy is
given by 
\begin{eqnarray}
F = -\frac{(2 m)^{3/2}}{15 \pi^2}
\left[
\left(
\mu - m - \alpha B
\right)^{5/2} 
+
\left(
\mu - m + \alpha B
\right)^{5/2}
\right]
+ \frac12 B^2 \, .
\label{freetotal}
\end{eqnarray}
In the following of this section, we analyze this free energy and study the ferromagnetism.


\subsection{Ferromagnetism at higher density}

\subsubsection{Complete polarization of the spins}

\begin{figure}
\begin{center}
\includegraphics[width=0.8\textwidth]{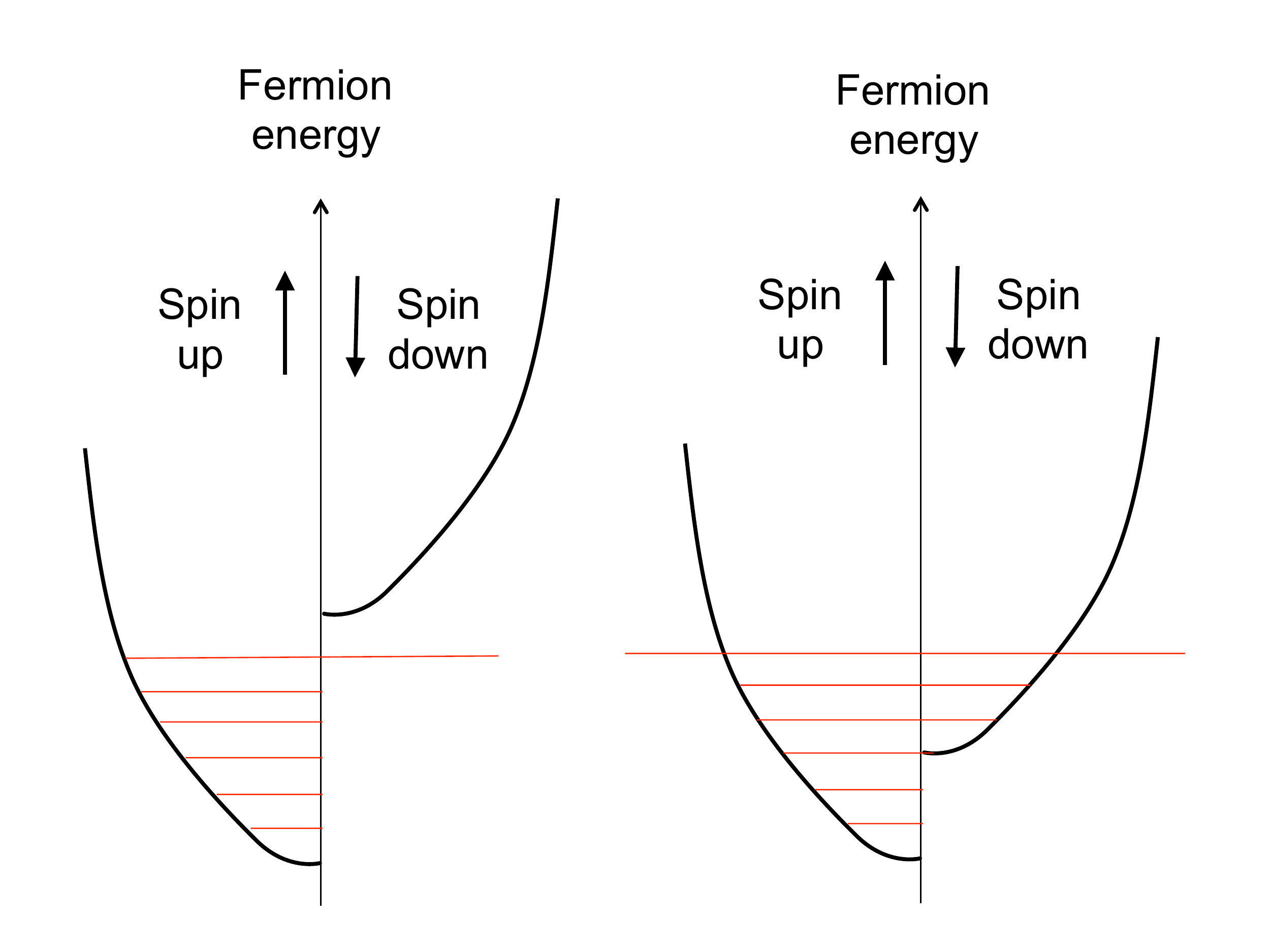}
\caption{The fermi surface and polarization of spins. The spin-magnetic coupling modifies the depth of the dispersion relation according to the fermion spins.
Left: all the spins are polarized. Right: there remains some density of 
the opposite component of the spin.}
\label{figspin}
\end{center}
\end{figure}

A large magnetic field is expected to correspond to a high density of the
fermion.
For a large magnetic field, one of the two terms for the spins 
in the free energy becomes
ill-defined; the expression (\ref{freetotal}) is valid only when
\begin{eqnarray}
\mu - m - |\alpha|B > 0 \, .
\end{eqnarray}
For a large magnetic field, 
this condition is not met. In that case,
we need to use the following expression for the free energy
\begin{eqnarray}
F = -\frac{(2 m)^{3/2}}{15 \pi^2}
\left(
\mu - m + |\alpha|B
\right)^{5/2}
+ \frac12 B^2 \, .
\label{freetotal2}
\end{eqnarray}
The spins are fully aligned (see Figure \ref{figspin} Left).

To turn this free energy (as a function of the chemical potential) to the energy
(as a function of the fermion density), let us make a Legendre transform.
The fermion number density is given by
\begin{eqnarray}
\rho \equiv - \frac{\partial F}{\partial \mu}
=\frac{(2 m)^{3/2}}{6 \pi^2}
\left(
\mu - m +  |\alpha|B
\right)^{3/2}\, .
\label{densfer}
\end{eqnarray}
Then the energy is given by
\begin{eqnarray}
E \equiv F + \mu \rho
= m \rho + \frac{3^{5/3}\pi^{4/3}}{2^{1/3} 5}\frac{1}{m} \rho^{5/3}
  - |\alpha| B \rho + 
\frac12 B^2 \, .
\end{eqnarray}
The interpretation of each term is quite clear.
The first term is the fermion mass energy, as $\rho$ is the number density of
the fermion. The second term is the Fermi energy. The third term is due to the
spin magnetic coupling. And the last term is for the magnetic field self energy.

We would like to find an energy minimum for a given fermion density $\rho$.
It is quite straightforward, since the last two terms in the energy can be written
as a perfect squared,
\begin{eqnarray}
E 
= m \rho + \frac{3^{5/3}\pi^{4/3}}{2^{1/3} 5}\frac{1}{m} \rho^{5/3}
+ \frac12 \left(B
 - |\alpha| \rho 
\right)^2
- \frac12 \alpha^2 \rho^2
\, .
\end{eqnarray}
So, to minimize the energy, a spontaneous magnetization should take place,
\begin{eqnarray}
B = |\alpha| \rho \, ,
\label{Bsol}
\end{eqnarray}
at which the energy density is given by
\begin{eqnarray}
E 
= m \rho + \frac{3^{5/3}\pi^{4/3}}{2^{1/3} 5}\frac{1}{m} \rho^{5/3}
- \frac12 \alpha^2 \rho^2
\, .
\label{enefreeres}
\end{eqnarray}


\subsubsection{Co-existence of both spins}

The magnetization at the high density in the description above assumes the complete polarization of
the fermions. At not-so-high density of the fermions, we expect that not all the fermions
are polarized (see Figure \ref{figspin} Right). Let us see indeed this is the case.

From the original total free energy (\ref{freetotal}), the equilibrium condition $\partial F / \partial B=0$ is
\begin{eqnarray}
0= 
-\frac{(2 m)^{3/2}}{15 \pi^2}
\frac{5}{2}\alpha
\left[-
\left(
\mu - m - \alpha B
\right)^{3/2} 
+
\left(
\mu - m + \alpha B
\right)^{3/2}
\right]
+ B \, .
\label{equifB}
\end{eqnarray} 
Note that this can be always satisfied at $B=0$. Therefore, no magnetization is
always a possibility of the equilibrium, and we need to compare if magnetized phase has
a lower energy density to conclude the ferromagnetism.
As we shall see, for lower density there is no magnetization, while for a high density
the ferromagnetism is preferred.

To see in more detail the density dependence, we calculate the density $\rho$ as
\begin{eqnarray}
\rho \equiv -\frac{\partial F}{\partial \mu}
= 
\frac{(2 m)^{3/2}}{6 \pi^2}
\left[
\left(
\mu - m - \alpha B
\right)^{3/2}
+\left(
\mu - m +  \alpha B
\right)^{3/2}
\right]
\, .
\end{eqnarray} 
We can eliminate $\mu$ by using the equilibrium condition (\ref{equifB}), to obtain
the equilibrium condition in terms of the density,
\begin{eqnarray}
\left(\rho + \frac{B}{\alpha}\right)^{2/3}
-\left(\rho - \frac{B}{\alpha}\right)^{2/3}
= \frac{4}{(3 \pi)^{2/3}} m \alpha B.
\label{rhoBfreeun}
\end{eqnarray} 
This equation determines the magnitude of the spontaneous magnetic field $B$, once 
the density $\rho$ is given. (Again, $B=0$ is an alternative solution satisfying this equation.)

We notice here that at a density
\begin{eqnarray}
\rho = \frac{1}{|\alpha|}B \, ,
\label{rhoBeq}
\end{eqnarray} 
the equation (\ref{rhoBfreeun}) can make one term vanish. This is nothing but the
point when we make a transition to the fully-polarized phase which we considered earlier.
Substituting (\ref{rhoBeq}) into (\ref{rhoBfreeun}), we obtain the threshold density
\begin{eqnarray}
\rho_2 = 
\frac{3^2 \pi^4}{2^4}\frac{1}{m^3\alpha^6}
\, .
\label{rho2}
\end{eqnarray} 
If the density is above this value, $\rho > \rho_2$, the system in fully polarized and
the analysis reduces to what we have considered earlier.

There is another condition for which the equation (\ref{rhoBfreeun}) can have
 a non vanishing solution for $B$. Using the following expansion
\begin{eqnarray}
(1 + \epsilon)^{2/3} - (1 - \epsilon)^{2/3}
= \frac43 \epsilon + \frac8{81} \epsilon^3 + {\cal O}(\epsilon^5) \, ,
\end{eqnarray} 
we notice that (\ref{rhoBfreeun}) can have a solution only when 
the slope around $B\sim 0$ can satisfy the following inequality
\begin{eqnarray}
\rho^{2/3} \cdot \frac43 \frac{1}{|\alpha|\rho} < \frac{4m|\alpha|}{(3\pi^2)^{2/3}} \, .
\end{eqnarray} 
This condition is rephrased as
\begin{eqnarray}
\rho_1 \equiv \frac{\pi^4}{3}
\frac{1}{m^2\alpha^6}
< \rho \, .
\end{eqnarray} 
When $\rho \leq \rho_1$, we find no solution for (\ref{rhoBfreeun}), other than $B=0$.
So, as is expected, for low density there is no ferromagnetic phase.

In summary, we find the following possible phases in our system;
\begin{eqnarray}
\left\{
\begin{array}{cccc}
&\rho \leq \rho_1 &     :  & \hspace{5mm}     B=0\nonumber \\
&\rho_1 < \rho < \rho_2 & : &  \hspace{5mm} B=0 \quad {\rm or}\quad 
B=\mbox{nontrivial solution of (\ref{rhoBfreeun}), spin mixed}
\nonumber \\
&\rho_2 \leq \rho & : &
\hspace{5mm} B=0 \quad {\rm or}\quad 
B=|\alpha| \rho \quad
\mbox{(spins fully polarized)}
\end{array}
\right.
\end{eqnarray}


\subsection{Favoring ferromagnetic phase}

To study whether this ferromagnetic order can actually occur in the system of our concern, let us compare the resultant energy (\ref{enefreeres}) with the energy with no magnetic field
(no magnetization). 

Putting $B=0$ reduces the system to that of the ordinary free fermion,
and in the non relativistic case, once given the density $\rho$, we know the total energy
\begin{eqnarray}
E_{B=0} = m \rho + \frac{3^{5/3} \pi^{4/3}}{10 m} \rho^{5/3} \, . 
\end{eqnarray} 
The first term is the energy contribution from the fermion mass, and the second term is the 
fermion kinetic energy integrated to the Fermi surface.

We compare this $E_{B=0}$ with the total energy density with the fully polarized spins (\ref{enefreeres}),
to have
\begin{eqnarray}
E - E_{B=0} = 
\frac{3^{5/3} \pi^{4/3}}{10m}(2^{2/3}-1) \rho^{5/3}
- \frac{1}{2} \alpha^2 \rho^2 \, .
\end{eqnarray} 
It is easy to show that this is always negative for the density $\rho \geq \rho_2$ which 
is the condition for the spin full polarization, see (\ref{rho2}). So, we conclude that indeed the 
ferromagnetic phase is preferred at the high density $\rho \geq \rho_2$.

It is also straightforward to show that even in the range $\rho_1 < \rho < \rho_2$, the ferromagnetic phase
$B\neq 0$ is preferred. To show this, 
we need numerical calculations since the energy for this phase is not expressed 
in an analytic form.

Finally, let us see the value of the chemical potential corresponding to the ferromagnetic phase, to find some consistency conditions; first, a thermodynamic stability condition, and second, the validity of the non-relativistic approximation. 
When all the spins are polarized,
we have (\ref{Bsol}) which can be substituted to the relation between the density and the chemical potential (\ref{densfer}), to find
\begin{eqnarray}
\mu = m + \frac{(6 \pi^2)^{2/3}}{2m} \rho^{2/3} - \alpha^2 \rho \, .
\label{rhomu}
\end{eqnarray} 
The thermodynamics stability condition is
\begin{eqnarray}
\frac{\partial \mu}{\partial \rho} > 0
\end{eqnarray} 
which tells just the fact that larger chemical potential provides a higher density.
Using our relation (\ref{rhomu}) at the high density ferromagnetic phase, we have
the thermodynamic stability condition
\begin{eqnarray}
\rho < \rho_3 \equiv \frac{2^2 \pi^4}{3}
\frac{1}{m^3 \alpha^6} \, .
\end{eqnarray} 
The value $\rho_3$ is larger than $\rho_2$, so the ferromagnetic phase is
stable for $\rho_2 < \rho < \rho_3$.

Second, we check the non-relativistic approximation. If we substitute 
the typical value $\rho = \rho_2$ for the ferromagnetic phase to the relation
(\ref{rhomu}), we find
\begin{eqnarray}
\mu-m+|\alpha| B = 2\cdot 3^2 \pi^4 \frac{1}{m \alpha^2} \, .
\end{eqnarray} 
This is the hight of the Fermi sea as measured from the bottom of the dispersion relation, so the non relativistic approximation is valid when this value is much smaller than the mass
$m$,
\begin{eqnarray}
\frac{2\cdot3^2 \pi^4}{m^2} \ll \alpha^2 \, .
\label{alphacons}
\end{eqnarray} 
So, our non relativistic approximation is valid when this condition is met
for the spin-spin interaction coefficient $\alpha^2$.

To gain more insight on the relation (\ref{alphacons}), let us adopt
hypothetically the expression of the magnetic moment for 
a charged fermion (\ref{alpha}) (although our fermion is neutral). Using (\ref{alpha}), 
the relation (\ref{alphacons}) is written as
\begin{eqnarray}
2^{5/2} 3 \pi^2 \ll |g| e \, .
\end{eqnarray} 
For example, 
the observed values for electrons are 
$|g|\sim 2$ and $e^2/4\pi \sim 1/137$, so this non-relativistic 
condition is not met.
Note however that in this paper we are interested in 
a neutral fermions, not the electron which
has a minimal coupling to the magnetic field.
In the next section, we
study neutrons in more details. 
We will find that, although the $g$-factor for the neutrons is not so large,
the non-relativistic approximation is valid:
in addition to the magnetic field coupling, there appears a pion coupling which
plays the same role, and the approximation is valid for the total interactions.
The pion condensation is the main subject of the next section.

\subsection{Similarity to Nambu-Jona-Lasinio model}

In the previous subsections, we have seen that the ferromagnetic phase is preferred
compared to the free neutral fermions, when the density is large enough.
Let us briefly discuss the reason why the simple model (\ref{lagful}) is expected to favor a ferromagnetic phase as for homogeneous phases. 
Indeed, we find an interesting relation to the famous Nambu-Jona-Lasinio (NJL) model
\cite{Nambu:1961tp,Nambu:1961fr}
in the following.
We can naturally assume that the phases under consideration is spatially homogeneous, 
therefore there is no electric field generated. In that case, the field $B_i$ serves as 
an auxiliary field  and we can integrate it out in our system 
(\ref{lagful}).\footnote{ 
Note that this integration is not allowed normally, but 
here we ignore the electromagnetic propagation. 
However, for a discussion of only a homogeneous phase, one can make the integration
and it provides an
intuitive picture.}
The resultant 
Lagrangian is
\begin{eqnarray}
{\cal L}_{\rm fermion} &= &\bar{\psi} \left(
i \partial_\mu \gamma^\mu - m + i\gamma_0 \mu 
\right)\psi 
+\frac12 \alpha^2  \left(\bar{\psi} \gamma_i \gamma_5 \psi\right)^2\, .
\label{spinspin}
\end{eqnarray}
Immediately we can see a resemblance to the NJL model, 
the renowned model for a spontaneous 
chiral symmetry breaking. The NJL model is characterized by 
a four-fermion interaction $(\bar{\psi}\psi)^2$, 
which can be
thought of as a squared of chiral condensate $\bar{\psi}\psi$. The four-fermion interaction
governs the condensation of the operator $\bar{\psi}\psi$. 
Our model can be considered as a generalization of the NJL model by replacing the
$(\bar{\psi}\psi)^2$ coupling with the spin-spin interaction\footnote{To recover the Lorentz invariance of the interaction term of the system, 
one can add the axial density squared term 
$\left(\bar{\psi} \gamma_0 \gamma_5 \psi\right)^2$ so that the interaction recovers the Lorentz invariance. It is not our scope of this paper.
}
 ${\cal S}_i {\cal S}_i$.
Since fermions possess spins, once we turn on a nonzero density for the fermions,
the spin-spin interaction may cause
a spontaneous magnetization, as in the case of the NJL model.
In fact, the spin-spin interaction is a popular interaction in condensed matter physics.
When the coefficient $\alpha^2$ of the last term is positive, the system is expected to
favor a spontaneous magnetization, {\it i.e.} a ferromagnetic phase. 

For the phase to
be realized, a high density would be necessary so that the neighboring fermions
can interact.\footnote{Note that in the NJL model, in contrast,  the fermion density is not necessary for the condensation, and the phase is unique.}
Therefore we also expect a phase transition from the normal phase to the
ferromagnetic phase as we increase the density, and the critical density should be a function of the 
coupling $\alpha$ and the mass of the fermion $m$ since these are the only parameters of
our system. This is what we have seen in this section, and the similarity to the NJL model allows us to intuitively understand the origin of the ferromagnetism.

Before ending this section, we should note one thing. 
Our analysis in this section assumes the homogeneity in space.
Normally one can allow inhomogeneous profile of the matter, which results in a
spontaneous emergence of a spatial modulation. 
A modulated phase would have smaller energy density compared to the ferromagnetic phase studied in this section. 
In the analysis in this section, we treated only a constant magnetic field $B_i$.  
However, normally the integration of $B_i$ as a constant
auxiliary field is not allowed, because photons propagate and $B_i$ is a part of
the photon kinetic term. Once one integrates out the electromagnetic field properly,
one finds a non-local action of fermions. The integrated nonlocal action can be used
for analyses of inhomogeneous phases of the fermions, see \cite{Maeda} for example.
In this paper we consider a homogeneous ferromagnetic phase, and whether it is
realized or not should be determined by a comparison with inhomogeneous phases.
As for the QCD application, we shall discuss this problem later in the next section.


\section{Chiral model of neutrons with pion condensation, magnetic field and anomaly}

We saw in the previous section that a generic 
neutral fermion system, with the simple Zeeman coupling, is shown to exhibit 
a ferromagnetism, under the assumption of the spatial homogeneity. 
As a concrete example, in this section we investigate a
neutron matter at a high density. Neutrons interact with each other not only via 
the magnetic field and the spin-magnetic interaction but also a pion exchange.
Interestingly, the two interactions have the same structure, under a simple profile
for a pion condensation. 
The pion condensation part is {\it a la} Dautry and Nyman \cite{DN}. 
In addition, QCD has an axial anomaly term which
relates the two condensations --- the magnetic field and the pion condensation,
and in fact enhances each other. The enhancement makes the total free energy 
decrease. We evaluate the total energy density of the ferromagnetic phase.
Finally we compare the resultant ferromagnetic phase with the
well-studied ALS (Alternating layer structure) phase for pion condensation.

\subsection{Dense neutrons and pions with axial anomaly}

\subsubsection{Axial anomaly for the pion Lagrangian}

Low energy action of QCD is given by the standard Lagrangian of the linear sigma model 
dictated by the breaking of the chiral symmetry,
\begin{eqnarray}
{\cal L} &= &\bar{\psi} \left(
i \partial_\mu \gamma^\mu - g(\sigma + i \gamma_5\mbox{\boldmath$\tau$}\cdot 
\mbox{\boldmath$\pi$})
\right)\psi 
\nonumber \\
& &
+ \frac{1}{2}(\partial_\mu \sigma)^2 + \frac{1}{2}(\partial_\mu \mbox{\boldmath $\pi$})^2
- m_\pi^2 f_\pi \sigma - V(\sigma^2\! +\! \mbox{\boldmath $\pi$}^2).
\end{eqnarray}
Here $\psi = (p,n)^{\rm T}$ is the nucleon field, and $\sigma$ and $\mbox{\boldmath $\pi$}$ are sigma model
fields leading to pions.
$f_\pi$ is the pion decay constant, and $m_\pi$ is the pion mass.
The global symmetry is the chiral symmetry $U(2)_{\rm L} \times U(2)_{\rm R}$. 
The chiral symmetry is broken due to the chiral condensate,
$\sigma^2 + \mbox{\boldmath $\pi$}^2= f_\pi^2$, which is realized by the potential term $V$.
Once the sigma model field obtains the expectation value, the nucleons acquire a mass, $g f_\pi = M_{\rm N}$. 
In this paper we do not consider the difference
of the masses for protons and neutrons. 

In the ideal case with no proton, and no charged pions, the Lagrangian is 
\begin{eqnarray}
{\cal L}_\sigma &= &\bar{\psi_n} \left(
i \partial_\mu \gamma^\mu - g(\sigma - i \gamma_5 \pi_3)
\right)\psi_n 
\nonumber \\
& &
+ \frac{1}{2}(\partial_\mu \sigma)^2 + \frac{1}{2}(\partial_\mu \pi_3)^2
- m_\pi^2 f_\pi \sigma - V(\sigma^2\! +\! \pi_3^2) \, ,
\end{eqnarray}
where $\psi_n$ is the neutron field. Since we want to deal with finite density of neutrons, we include a 
chemical potential term for the neutron,
\begin{eqnarray}
{\cal L}_n = i \bar{\psi}_n \gamma_0 \mu_n \psi_n \, . 
\end{eqnarray}
In the presence of the magnetic field with which the neutrons interact through their magnetic moment,
we add  the following Lagrangian,
\begin{eqnarray}
{\cal L}_B = -\frac{1}{2} B_i^2 + \frac12 \bar{\psi}_n \gamma_i \gamma_5 \psi_n \frac{g_n e}{2 M_{\rm N}}B_i
\, .
\end{eqnarray}
The first term is the energy of the magnetic field. The second term is the Pauli term for the interaction between the magnetic moment of the neutron and the magnetic field. Note that the spin density of the fermion is given by
$\frac12 \bar{\psi}_n \gamma_i \gamma_5 \psi_n$, and $g_n$ is the neutron $g$ factor. 

In the presence of the magnetic field and the neutral pion condensation which is spatially dependent,
there  exists an axial anomaly term,
\begin{eqnarray}
{\cal L}_{\rm anom} =-i  \frac{e}{4\pi^2 f_\pi^2} \mu_{\rm em}  
\biggl[(\sigma + i \pi_3)^\dagger \partial_i (\sigma + i \pi_3)\biggr]
B_i\, .
\end{eqnarray}
Here $\mu_{\rm em}$ is the electromagnetic chemical potential. This term is relevant for, for example,
the neutral pion decay $\pi^0 \rightarrow 2\gamma$ via the axial anomaly, as is seen from the fact that the electromagnetic chemical potential can be thought of as a constant background electrostatic potential $A_0^{\rm (em)}$. 
So, our total Lagrangian is
\begin{eqnarray}
{\cal L} = {\cal L}_\sigma + {\cal L}_n + {\cal L}_B + {\cal L}_{\rm anom}.
\label{totalLag}
\end{eqnarray}

\subsubsection{Neutral pion condensation}

We consider a neutral pion condensate, following Dautry and Nyman \cite{DN}, 
\begin{eqnarray}
\sigma + i \pi_3 = f_\pi \exp (i {\bf q}\cdot {\bf x}) \, , 
\quad
\pi_1 + i \pi_2 =0 \, .
\label{DN}
\end{eqnarray}
This corresponds to a specific condensation of the neutral pion in the nonlinear representation, since the relation between the linear and non-linear representation is
$\sigma + i\pi_3 \sim f_\pi  \exp (i \Pi_0(x))$ where $\Pi_0(x)$ is the physical neutral pion excitation. The condensation (\ref{DN}) corresponds to 
$\Pi_0(x) \sim {\bf q}\cdot {\bf x}$, a linear profile in space.
This can be regarded as a dense parallel domain walls which was considered in the context of anomaly-enhanced pion condensation  in
\cite{Eto:2012qd}. 

In this paper, we shall generalize the study of Dautry and Neyman (\ref{DN}), to
include the magnetization and the QCD anomaly.
With this condensation (\ref{DN}), 
the anomaly term ${\cal L}_{\rm a nom}$ is given simply as
\begin{eqnarray}
{\cal L}_{\rm anom} = \frac{e}{4\pi^2} \mu_{\rm em} q_i B_i\, .
\end{eqnarray}
In the following, without losing generality, we can turn on only the $x^3$ components of the
magnetic field and $q_3$, which will be denoted as $B$ and $q$.

According to Dautry and Nyman, if the condensation $q$ is large, the neutron spins are fully polarized.
In the non-relativistic approximation for the neutron Fermi momentum, the free energy for the 
free neutrons in the background pion condensation and the magnetic field is derived from the Lagrangian above,
\begin{eqnarray}
F_n = -\frac{(2 M_{\rm N})^{3/2}}{15 \pi^2}
\left(
\mu_n - M_{\rm N} + \frac12 g_A q - \frac{g_n e B}{4 M_{\rm N}}
\right)^{5/2}.
\label{freen}
\end{eqnarray}
Here we have introduced the axial coupling $g_A$ which is, at the tree level, equal to $g$ in the 
$\sigma$-model Lagrangian ${\cal L}_\sigma$.
The total Free energy including the pion condensation and the magnetic field is
\begin{eqnarray}
F = F_n + f_\pi^2 m_\pi^2 + \frac12 f_\pi^2 q^2 + \frac12 B^2 - \frac{e}{4\pi^2} \mu_{\rm em} q B.
\label{free}
\end{eqnarray}
The second term is from the pion mass term together with the pion condensation (\ref{DN}). The third term
is from the pion kinetic term with (\ref{DN}). The last term is the axial anomaly term.

\subsubsection{Hamiltonian and the neutron density carried by the anomaly}

Our interest is the core of the neutron star where we have the $\beta$-equilibrium. 
In addition to the neutrons, there exist protons and electrons.
The electromagnetic chemical potential is given by
\begin{eqnarray}
\mu_{\rm em} = \frac12 (\mu_p-\mu_e)
\end{eqnarray}
where $\mu_p$ and $\mu_e$ are proton and electron chemical potential, respectively.\footnote{The factor $1/2$ should be there, because the total Free energy given with the number density should be 
$\mu_p \rho_p + \mu_e \rho_e$, and the total electric charge is $\rho_p-\rho_e$.  Its canonical conjugate is
$\frac12 (\mu_p-\mu_e)$.}
Assuming the $\beta$-equilibrium, we impose
$\mu_n = \mu_p + \mu_e$.
And since we approximate the system by the pure neutron matter for simplicity, we also impose
the charge neutrality condition in a trivial manner, $\rho_p = \rho_e=0$, which is equivalent to have
$\mu_p = M_{\rm N}$. Then the anomaly term in the free energy (\ref{free}) is written as
\begin{eqnarray}
- \frac{e}{4\pi^2} \mu_{\rm em} q B = - \frac{e}{4\pi^2} \frac12 (\mu_p - (\mu_n-\mu_p)) q B
=- \frac{e}{4\pi^2} \left(M_{\rm N} - \frac12\mu_n\!\right) q B
\, .
\end{eqnarray}

We evaluate the energy density of the system as a function of the neutron density and
the condensation $q$ and the magnetic field $B$. The neutron density is computed as
\begin{eqnarray}
\rho_n = -\frac{\partial F}{\partial \mu_n}
=\frac{(2 M_{\rm N})^{3/2}}{6 \pi^2}
\left(
\mu_n - M_{\rm N} + \frac12 g_A q - \frac{g_n e B}{4 M_{\rm N}}
\right)^{3/2}
+ \frac{e}{8\pi^2} q B \, .
\label{dens}
\end{eqnarray}
The last term is the anomaly-induced baryon charge. Using the expression, the final result for the energy density is given by
\begin{eqnarray}
E &=& F + \mu_n \rho_n \nonumber \\
&=& \frac{3^{5/3}\pi^{4/3}}{2^{1/3} 5}\frac{1}{M_{\rm N}} \left(\rho_n - \frac{e}{8 \pi^2}q B\right)^{5/3}
+ \left(M_{\rm N} - \frac12 g_A q + \frac{g_n e B}{4 M_{\rm N}}\right)\left(\rho_n - \frac{e}{8 \pi^2}q B\right)
\nonumber \\
& & + f_\pi^2 m_\pi^2 + \frac12 f_\pi^2 q^2 + \frac12 B^2 - \frac{e}{4\pi^2} M_{\rm N} q B \, .
\label{ene}
\end{eqnarray}

For a comparison, we write the expression of Dautry and Nyman \cite{DN}:
\begin{eqnarray}
E &=& \frac{3^{5/3}\pi^{4/3}}{2^{1/3} 5}\frac{1}{M_{\rm N}} \left(\rho_n\right)^{5/3}
+ \left(M_{\rm N} - \frac12 g_A q \right)\rho_n
\nonumber \\
& & + f_\pi^2 m_\pi^2 + \frac12 f_\pi^2 q^2.
\label{eneDN}
\end{eqnarray}
This is obtained from (\ref{ene}) by just putting $B=0$. The difference from 
just the pion condensation is obvious.
Let us look at the second term of (\ref{end}), which in fact exhibits the nature of our model explicitly.
In the absence of the pion condensation and the magnetic field, the second term is simply 
$\rho_n M_{\rm N}$. This is the cost of the energy due to the mass of the neutron.
Now, the cost for each neutron can be reduced by the pion condensation due to the axial coupling, and 
by the magnetic field times the neutron magnetic moment, as 
$M_{\rm N} \rightarrow M_{\rm N} - \frac12 g_A q + \frac{g_n e B}{4 M_{\rm N}}$.
Furthermore, the anomaly term can reduce effectively the density of the neutrons,
$\rho_n \rightarrow \rho_n - \frac{e}{8 \pi^2}q B$. In addition, the last term of (\ref{ene}) is for the QCD anomaly, and it makes the total energy decrease further.

\subsection{Spontaneous magnetization and the pion condensation}

For a given density $\rho_n$ of the neutrons, we can minimize the energy $E$ (\ref{ene}). Later we present 
our numerical results.  But here, to explain the intrinsic behavior of the system, we evaluate the minimization of the energy in
the absence of the anomaly term. 
Without the anomaly term, the energy density is simplified as
\begin{eqnarray}
E &=& \frac{3^{5/3}\pi^{4/3}}{2^{1/3} 5}\frac{1}{M_{\rm N}} \rho_n^{5/3}
+\!  \left(\! M_{\rm N}  -  \frac12 g_A q + \frac{g_n e B}{4 M_{\rm N}}\right)\rho_n
\nonumber \\
& &
+ f_\pi^2 m_\pi^2 + \frac12 f_\pi^2 q^2 + \frac12 B^2 . \;\;
\label{ene2}
\end{eqnarray}
The energy is quadratic in $q$ and $B$, so we can analytically find the minimum of the
energy. In fact, the energy density expression (\ref{ene2}) 
can be brought to the following  form with the
perfect squared,
\begin{eqnarray}
E  &=& E_0 
+ \frac12 f_\pi^2 
\left(
q 
- \frac{g_A}{2 f_\pi^2} \rho_n
\right)^2
+ \frac12 
\left(
B + 
\frac{g_ne}{4 M_{\rm N}} 
\rho_n
\right)^2\, , \;\;
\end{eqnarray}
where the minimum energy is
\begin{eqnarray}
E_0 &=& \frac{3^{5/3}\pi^{4/3}}{2^{1/3} 5}\frac{1}{M_{\rm N}} \rho_n^{5/3}
+ M_{\rm N}\rho_n
-\frac{g_A^2}{8 f_\pi^2} \rho_n^2
-
\frac{g_n^2 e^2}{32 M_{\rm N}^2} 
\rho_n^2 \, .
\label{sq}
\end{eqnarray}
The last term in the minimum energy density $E_0$ is due to the
magnetic field. The first three terms are that of Dautry and Neyman \cite{DN},
and compared to that, our energy is smaller by the last term.

The minimization of the energy is achieved when the perfect squares in (\ref{sq}) vanish,
\begin{eqnarray}
B &=& \left(
\frac{-g_n}{4 M_{\rm N}} 
\right) 
e \rho_n \, ,
\label{B} 
\\
q &=&  
\frac{g_A}{2 f_\pi^2} \rho_n \, .
\label{q}
\end{eqnarray}
We have obtained the spontaneous magnetization of the neutron matter. The generated magnetic field is a monotonic function of the density, and in particular in this case of the absence of 
the anomaly, it is a linear function in 
the density.

The free energy $F_n$ for the neutrons, (\ref{freen}), is for fully polarized neutrons. Let us check if
this can be achieved for $q$ and $B$ which we obtained above for given $\rho_n$. 
The condition that 
the opposite spin state is absent is
\begin{eqnarray}
\mu_n - M_{\rm N} - \frac12 g_A q + \frac{g_n e B}{4 M_{\rm N}} < 0.
\end{eqnarray}
This means that the Fermi sea for the opposite spin state is below the conducting band.
At the minimum of the energy, we obtain $\rho_n$ dependence of the chemical potential $\mu_n$ 
from (\ref{dens}) with the solution (\ref{q}) and (\ref{B}) as
\begin{eqnarray}
\mu_n = \frac{(6 \pi^2)^{2/3}}{2 M_{\rm N}} \rho_n^{2/3}
+ M_{\rm N} 
-\left(
\frac{g_A^2}{4 f_\pi^2} + \frac{g_n^2 e^2}{16 M_{\rm N}^2} 
\right) \rho_n.
\end{eqnarray}
In terms of $\rho_n$, the condition of the full polarization is equivalent to
\begin{eqnarray}
(6 \pi^2 \rho_n)^{2/3} < 4 M_{\rm N}
\left[
\frac{g_A^2}{4 f_\pi^2}
+
\frac{g_n^2 e^2}{16 M_{\rm N}^2} \right]
\rho_n \, .
\end{eqnarray}
Substituting values as $M_{\rm N} = 938$ [MeV], $e^2/4\pi = 1/137$, $f_\pi=95$ [MeV], $g_A=1$, $g_n = -3.8$
and $m_\pi=135$ [MeV], we obtain 
\begin{eqnarray}
\rho_n > 0.39 \, [{\rm fm}^{-3}] \, .
\end{eqnarray}
This shows that for the density around twice of the standard nuclear density, all the spins are polarized.

Another constraint comes from a thermal stability condition. At any thermal equilibrium, we need to make sure
\begin{eqnarray}
\frac{\partial \mu_n}{\partial \rho_n} >0 \, .
\label{ther}
\end{eqnarray}
The condition (\ref{ther}) can be evaluated as
\begin{eqnarray}
\rho_n < 57.6 \, [{\rm fm}^{-3}] \, .
\end{eqnarray}
The bound is extremely high density and unrealistic, so this thermodynamic instability region is far above realistic
neutron density.

\begin{figure}
\begin{center}
\includegraphics[width=0.43\textwidth]{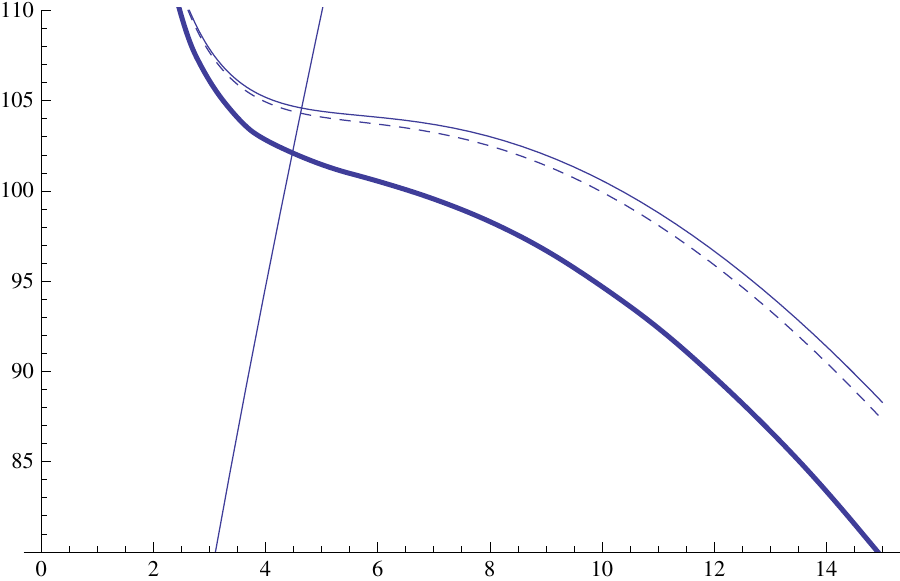}
\put(5,5){$\rho_n/\rho_0$}
\put(-255,130){$(E_0-M_{\rm N})/\rho_n$ [MeV]}
\caption{
A plot of the energy per a neutron, as a  function
of the neutron density $\rho_n$. Straight line: ordinary neutron matter
without the pion condensation. Thick curved line: our result with both
pion condensation $q$ and magnetization $B$ with the QCD anomaly term.
Thin curved line: the result of Dautry and Neyman \cite{DN} with only the
pion condensation $q$. Dashed line: the energy with both $q$ and $B$ but
without the QCD anomaly term.
}
\label{figano}
\end{center}
\end{figure}

\subsection{Anomaly enhancement and comparison to the ALS phase}
\label{section:ALS}

In the previous subsection, we found that even without the anomaly term the
total energy density is lowered by the magnetic field. In fact, the magnetic coupling
works in the same manner as the pion coupling.
Now, let us see how the anomaly term can help the condensation. 
The full expression for the total energy density including the anomaly term
was given in (\ref{ene}), and we can find the minimum energy configuration by
varying $q$ and $B$. 
Analytic analysis is not easy since the energy is not quadratic in $q$ and $B$, 
so we perform a numerical analysis to find the
energy minimum. 
The numerical results are summarized in Fig.~\ref{figano}.

Fig.~\ref{figano} is a plot of the energy per a neutron, as a  function
of the neutron density $\rho_n$. The neutron density is normalized by $\rho_0$ which is
the standard nuclear density. The thick line is our result with the anomaly term.
We observe that for a larger density, the energy per a neutron decreases.

In Fig.~\ref{figano}, for a comparison, 
we show a thin curved line which 
is the result of Dautry and Nyman \cite{DN} (that is, with no magnetic field $B$
but with the pion condensation $q$).
The dashed line in Fig.~\ref{figano} is the energy density with both $q$ and $B$ but without the anomaly term. We can see that the anomaly
term makes the energy per a neutron decrease. The straight line on the left is
for free neutrons without the pion condensation $q$ and without $B$.
So, as a comparison to the ordinary neutron matter, we see that the ferromagnetic phase
is preferred at high density.

\begin{figure}
\begin{center}
\includegraphics[width=0.43\textwidth]{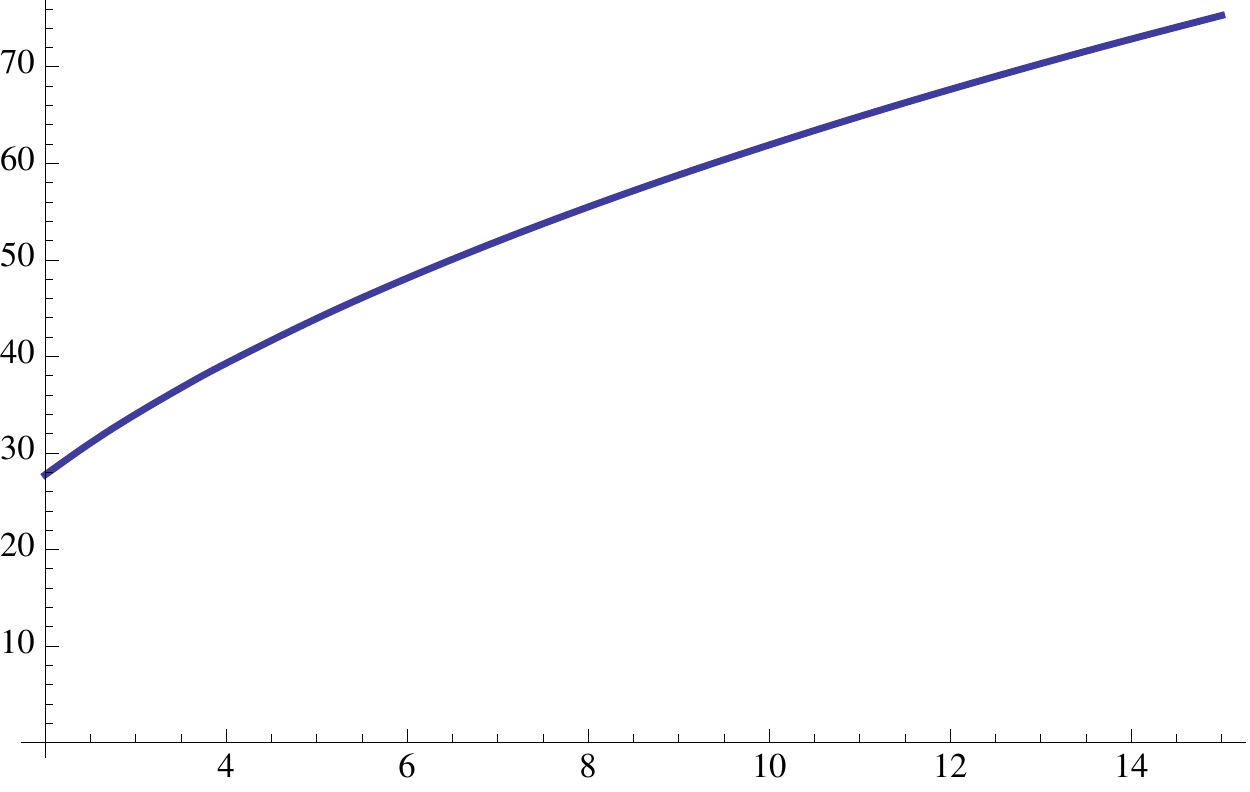}
\put(5,5){$\rho_n/\rho_0$}
\put(-205,130){$\sqrt{eB}$ [MeV]}
\caption{
A plot of the magnetic field spontaneously generated, as a  function
of the neutron density $\rho_n$. The neutron density is normalized by $\rho_0$ (the 
standard nuclear density). The evaluation is with the fully-polarized neutrons.}
\label{figB}
\end{center}
\end{figure}

We plot the magnetic field as a function of $\rho_n$, in Fig. \ref{figB}. It is a monotonic function of the neutron density. We find that the magnitude of the generated magnetic field is 
${\cal O}(10^2)$ [MeV] and thus it is of the QCD scale.
$\sqrt{eB}\sim 40$[MeV] corresponds to ${\cal O}(10^{17})$[G].

\begin{figure}
\begin{center}
\includegraphics[width=0.5\textwidth]{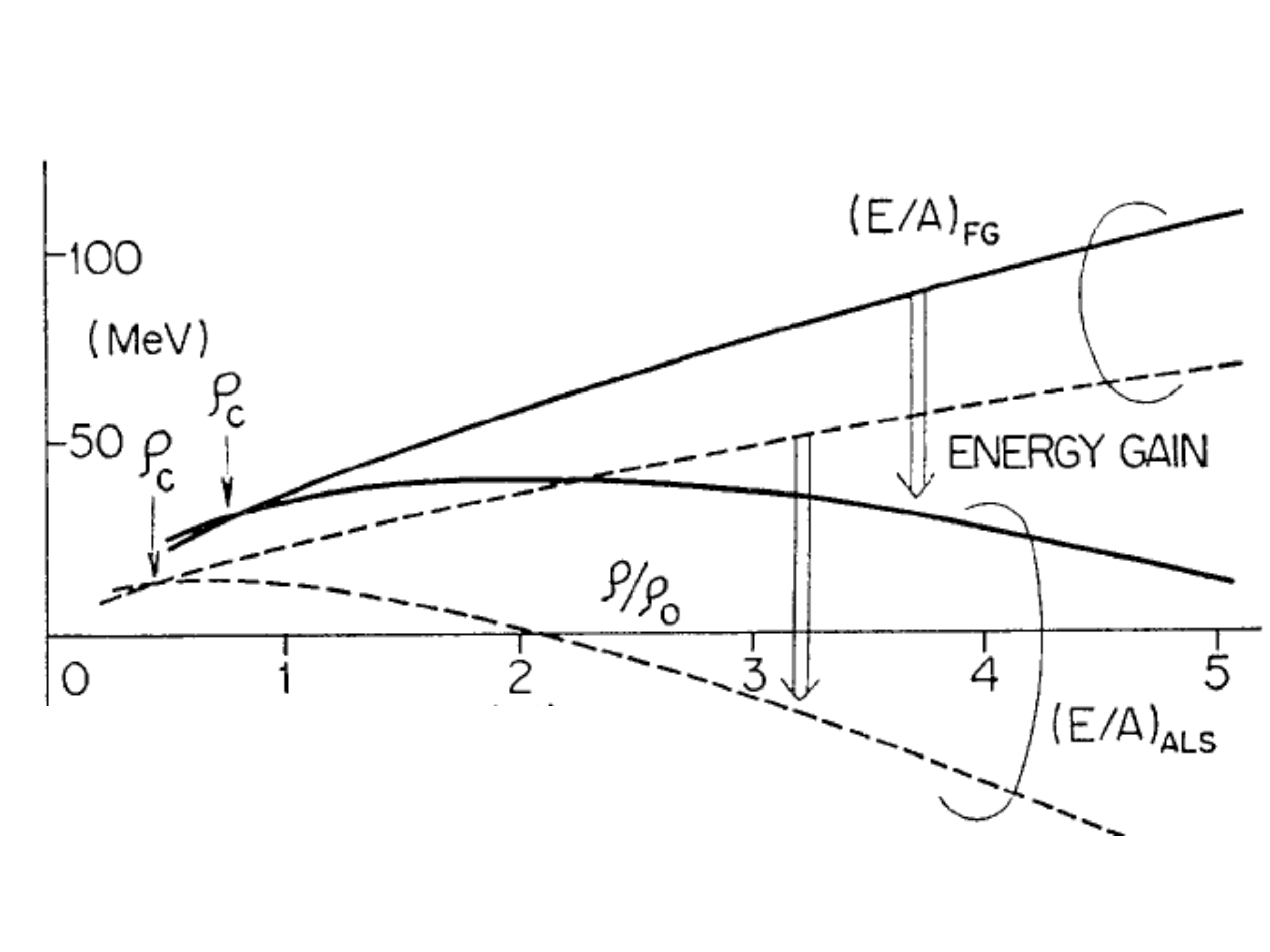}
\caption{An energy plot taken from \cite{Takatsuka:1978ku}, comparing the ALS phase
and the Fermi gas (ordinary neutron matter). The horizontal axis
is the neutron density  $\rho$ in the unit of the standard nuclear density $\rho_0$, 
while the vertical axis is the energy gain per a neutron. The upper curve is for the Fermi gas and the lower curve is for the ALS. Solid lines are for neutrons, and 
dashed lines are for symmetric nuclear matter. The arrows indicate the energy reduction by the ALS.
}
\label{figALS}
\end{center}
\end{figure}

Now, let us discuss whether our ferromagnetic phase is favored or not, in reality. 
The famous phase
for a pion condensation is the ALS (Alternating layer structure) phase 
\cite{Tamagaki:1976ds,Takatsuka:1977uj,Takatsuka:1978ku}, and we can compare
the result of the ALS phase with ours. 
See the result of the ALS (Fig.~\ref{figALS}) and compare it with our result (Fig.~\ref{figano}). 
Since already around $\rho_n/\rho_0 \sim 5$ the 
energy reduction 
of the ALS phase compared to the ordinary neutron matter is $70$ [MeV] (see Fig.~\ref{figALS}), while
for our ferromagnetic phase the energy reduction is only $10$ [MeV].
So, from this comparison, we conclude that the ALS phase is favored against
our ferromagnetic phase.

Our analysis in this paper is with the simplest model of neutrons, and we have not included
full nuclear forces. Once we include them in addition to our neutral pion coupling and 
the Zeeman coupling, the total free energy may change. In fact, in the 
following subsections, we include axial vector meson condensation and it
makes the total energy further decrease drastically.

In summary, here our observation
is that the ferromagnetism is closely related to the neutral pion condensation, and
the axial anomaly can help the total energy to decrease and enhance the magnetic field. The ferromagnetic phase has an energy density smaller than that of 
the ordinary neutron matter. But the energy of the ALS phase is smaller, in the approximation presented.

\subsection{Inclusion of axial vector meson condensation}
\label{section:axial}

In the previous subsections, we considered only the neutral pion field for the coupling to the
spins of the neutrons. The spin operator of the neutron is nothing but the spatial components
of the axial current, and 
in QCD, we expect infinite number of quark bound states which can couple to the 
axial current. They are axial vector mesons whose spectrum starts at the lowest with 
$a_1(1260)$ meson. In this subsection, we shall add the contribution of this lowest 
axial vector meson and will find that it will further make the free energy decrease, together
with the full spin polarization. 

The main system which we treat in this paper is described by (\ref{lagful}). On the other hand,
the effective Lagrangian for the axial vector meson $a_\mu(x)$ coupled to the axial current
and the neutrons is
\begin{eqnarray}
{\cal L} &= &\bar{\psi} \left(
i \partial_\mu \gamma^\mu - m + i\gamma_0 \mu 
\right)\psi 
\nonumber \\
&&
+\, \frac12 g_{aNN} \, \bar{\psi} \gamma_\mu \gamma_5 \psi \,a^\mu
+\frac14 \left(
\partial_\mu a_\nu - \partial_\nu a_\mu
\right)^2
-\frac{m_a^2}{2} a_\mu^2 \, .
\end{eqnarray}
Here, $g_{aNN}$ is the coupling of the axial vector meson to the neutron axial current,
and $m_a$ is the mass of the meson. For the lowest 
$a_1(1260)$ meson, the measured value 
is $m_a = 1230\pm 40$ [MeV].

As before, we concentrate on a homogeneous phase, and let us assume a constant
vacuum expectation value of the spatial component of the axial vector meson,
\begin{eqnarray}
\langle a_i \rangle = \mbox{const.}\neq 0 \, .
\end{eqnarray}
Then, re-writing $b_i \equiv a_i m_a$, the effective Lagrangian is now
\begin{eqnarray}
{\cal L} &= &\bar{\psi} \left(
i \partial_\mu \gamma^\mu - m + i\gamma_0 \mu 
\right)\psi 
+\frac{g_{aNN}}{2m_a} \bar{\psi} \gamma_i \gamma_5 \psi \,b_i
-\frac12 b_i^2 \, .
\end{eqnarray}
We observe that this Lagrangian has precisely the same form as (\ref{lagful}), so
the same mechanism of lowering the free energy by spin alignment can work.

This addition of the axial vector meson to the pion system 
modifies the total free energy (\ref{free})
a little bit. The resultant free energy is
\begin{eqnarray}
F &=& 
-\frac{(2 M_{\rm N})^{3/2}}{15 \pi^2}
\left(
\mu_n - M_{\rm N} + \frac12 g_A q - \frac{g_n e}{4 M_{\rm N}}B + \frac{g_{aNN}}{2m_a} b
\right)^{5/2}
\nonumber \\
&&
+ f_\pi^2 m_\pi^2 + \frac12 f_\pi^2 q^2 + \frac12 B^2 
+\frac12 b^2- \frac{e}{4\pi^2} \mu_{\rm em} q B.
\label{freeax}
\end{eqnarray}
Here we determined the orientation of $b_i$ in space such that it may strengthen the
spin polarization, and denote $b$ as the magnitude of $b_i$. Note that the axial vector meson
$b$ enters exactly in the same manner as that of the pion condensation $q$ and the
magnetic field $B$ except for the anomaly term (the last term in (\ref{freeax})).
So, basically the addition of the axial vector condensation enhances the
spin polarization of the neutrons, and further reduces the energy density.

As before, to gain an intuition of the behavior of the system, we first analyze the 
system without the anomaly term. Then the total energy is
\begin{eqnarray}
E &=& \frac{3^{5/3}\pi^{4/3}}{2^{1/3} 5}\frac{1}{M_{\rm N}} \rho_n^{5/3}
+  \left(\! M_{\rm N}  -  \frac12 g_A q + \frac{g_n e B}{4 M_{\rm N}}
+ \frac{g_{aNN}}{2m_a} b\right)\rho_n
\nonumber \\
& &
+ f_\pi^2 m_\pi^2 + \frac12 f_\pi^2 q^2 + \frac12 B^2 + \frac12 b^2 . \;\;
\label{ene3}
\end{eqnarray}
Compared to (\ref{ene2}), we find that we have additional terms
\begin{eqnarray}
\Delta E =  \frac{g_{aNN}}{2m_a} b\rho_n + \frac12 b^2 \, ,
\end{eqnarray}
which is independent of the other variables $q$ and $B$. So we can minimize it
independent of the other terms, and find
\begin{eqnarray}
\Delta E_0 = - \frac{g_{aNN}^2}{8m_a^2} \rho_n^2
\label{DelE0}
\end{eqnarray}
with an axial vector condensation
\begin{eqnarray}
|\langle a_i \rangle| = \frac{b}{m_a} = \frac{g_{aNN} }{2m_a^2}\rho_n \, .
\end{eqnarray}

To evaluate the energy $\Delta E_0$ in (\ref{DelE0}), we need the value of the 
axial vector coupling $g_{aNN}$. We refer to a generic argument of the chiral symmetry
by regarding the axial vector meson as a gauge boson of the symmetry 
\cite{Weinberg:1966fm,Schwinger:1967tc,Wess:1967jq},
\begin{eqnarray}
\frac{g_{aNN}}{m_a} = \frac{2g_A}{m_\pi}.
\label{chiralgann}
\end{eqnarray}
A naive substitution $g_A \sim 1$ and the mass for the pion and the $a_1$ meson
provides $g_{aNN} \sim 18$. 
Another estimate is as follows. We take care of one of the other equations coming from
the chiral symmetry argument \cite{Weinberg:1966fm,Schwinger:1967tc,Wess:1967jq}, 
$m_a = \sqrt{2} m_\rho$ which is not well satisfied by
the physical masses of the $\rho$ meson and the $a_1$ meson. So instead of using
the $a_1$ meson mass in the chiral symmetry formula (\ref{chiralgann}) we may use
the $\rho$ meson mass $m_\rho = 770$ [MeV]. Then we obtain $g_{aNN}\sim 16$.
However, a lattice simulation with an axial vector dominance provides
$g_{aNN}\sim 9$ (see for example \cite{Coon:1996wj}), 
so there is uncertainty for the coupling.

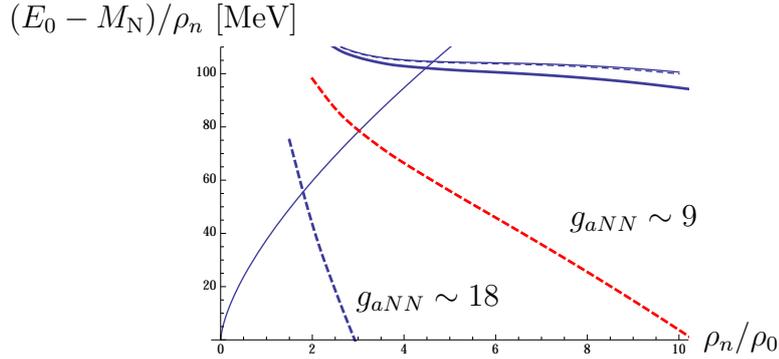
\begin{figure}
\begin{center}
\includegraphics[width=0.43\textwidth]{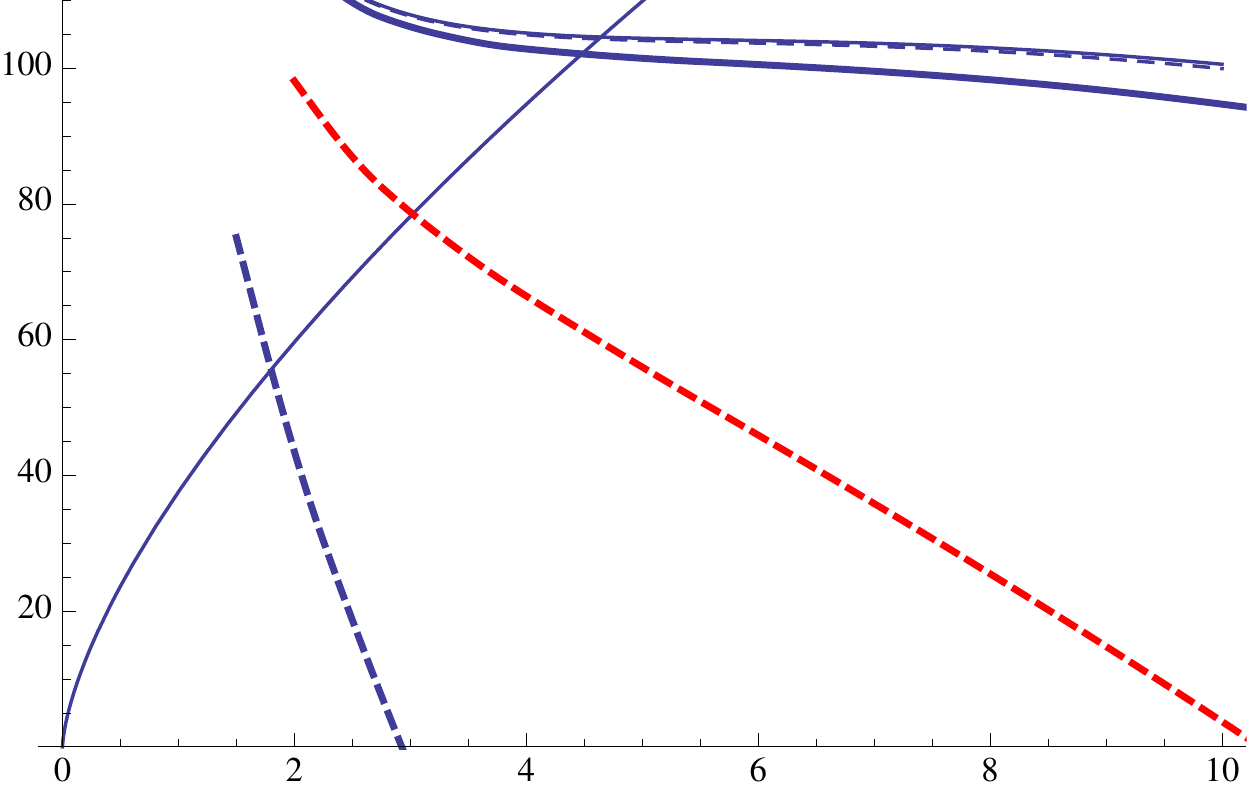}
\put(5,5){$\rho_n/\rho_0$}
\put(-45,50){$g_{aNN}\sim 9$}
\put(-125,20){$g_{aNN}\sim 18$}
\put(-255,125){$(E_0-M_{\rm N})/\rho_n$ [MeV]}
\caption{
A plot of the energy per a neutron, as a  function
of the neutron density $\rho_n$. On the previous figure for the pion condensation, here
we added two thick dashed lines, showing the axial vector condensation. 
The upper thick dashed line is for $g_{aNN}=9$ and the lower is for $g_{aNN}\sim 18$.}
\label{figano2}
\end{center}
\end{figure}

In our numerical estimate of the energy density,
we choose two typical values, $g_{aNN}\sim 18$ and $9$. 
Our result is shown in fig.~\ref{figano2}. We find that the energy per a nucleon drastically reduces further. 
Compared to the ALS phase, the case with $g_{aNN}\sim 18$ 
has a lower energy and thus favored. The case with $g_{aNN}\sim 9$ is almost
at the same order with the ALS phase.

\subsection{AdS/CFT treatment with a large $N_c$ approximation}
\label{section:axial2}

The AdS/CFT correspondence 
\cite{Maldacena:1997re,Gubser:1998bc,Witten:1998qj}
is a well-estabilished tool for analyzing strongly coupled gauge theory in a certain limit, and its application to QCD-like gauge theories were widely studied.
However the AdS/CFT tools for strongly coupled gauge theories work practically for
large $N_c$ gauge theories and at the limit of strong coupling, so
it would not be suitable
for precision analysis such as the energy gain via the condensation which is our interest in this paper. Nevertheless, it is important to find what kind of couplings among hadrons and the 
magnetic field is present in QCD, and what is the order of magnitude of the couplings.
The AdS/CFT approach, called holographic QCD, is suitable for that purpose, and 
in this short subsection we shall investigate it.

We use Sakai-Sugimoto model \cite{Sakai:2004cn,Sakai:2005yt} which is the stringy setup closest to QCD at present.
The nucleon meson couplings were obtained in \cite{Hashimoto:2008zw,Hashimoto:2009ys,Hong:2007kx,Hong:2007ay}, and the QCD anomaly term
was calculated in \cite{Sakai:2005yt}. 

Basically in holographic QCD we have a tower of mesons,
and this is true for the $a_1$ mesons. We have infinite number of axial vector mesons.
On the other hand, we have only a single pion (that is, in the model there does not appear
excited resonances of the pion). 

It is easy to read from \cite{Sakai:2005yt} 
that the axial vector mesons does not participate
in the QCD anomaly term, so the only contribution to the anomaly term is the pion
coupling which we considered in this paper. So we do not need to take care of 
all the mixing between the axial vector mesons and the magnetic field in the anomaly term,
at the leading large $N_c$ expansion and at the strong coupling limit.

On the other hand, the contribution of the axial vector meson to the nucleon spins, 
which we considered in the previous subsection, comes to a concern.
Since we have infinite number of axial vector mesons, all piles up as a sum and 
would cause possibly a tremendous contribution. We shall discuss the issue in the following.

First, in the AdS/CFT correspondence, 
the axial vector mesons are gauge fields at higher dimensions, and their interaction terms
are basically given by the Yang-Mills action in the higher dimensions.
We need to excite only the $\tau_3$ component of the isospin, while the Yang-Mills
action contains only a commutator-type interaction, so the direct interaction among
the constant axial vector mesons vanish.
This means that we does not need to consider the inter-level interaction of the axial vector 
meson tower.

We have seen in the previous subsection that a single axial vector meson reduces the
total energy by (\ref{DelE0}), so when there  exists a tower of the axial vector mesons
we have an energy reduction
\begin{eqnarray}
\Delta E_0 = - \rho_n^2 \sum_{i=1}^\infty 
r^{(i)}, \quad
r^{(i)} \equiv 
\left(
\frac{g_{a^{(i)}NN}}{8 m_{a^{(i)}}}
\right)^2
\end{eqnarray}
where $i$ is the label of the resonances, and $i=1$ corresponds to the lowest 
$a_1(1260)$. From this expression, we observe that all axial vector mesons contribute
additively, and the issue is the magnitude of the ratio $g_{a^{(i)}NN}/ m_{a^{(i)}}$ when $i$ 
increases.

The ratio can be calculated analytically by the AdS/CFT correspondence \cite{Hashimoto:2008zw}. 
However, the approximation of large $\lambda$ is not good, so
here we provide only the resulting numbers for a reference. The method developed in
\cite{Hashimoto:2008zw} can be generalized easily for higher axial vector mesons, and we find
\begin{eqnarray}
r^{(2)}/r^{(1)} \simeq 1.06,  \quad
r^{(3)}/r^{(1)} \simeq 1.07
\end{eqnarray}
at the large t'Hooft coupling limit. So the ratio does not decrease for larger $i$. 
This would be natural from the original idea of gauged chiral symmetry by Wess and Zumino
\cite{Wess:1967jq} which derived the relation (\ref{chiralgann}). 
Therefore, the
effect of the inclusion of the higher axial vector mesons is important, and it has an effect of further reducing the total energy density.

At large $N_c$ limit, all the axial vector meson tower reasonably contribute since the meson width is narrow, and one would imagine the tremendous amount of energy reduction by
introducing all the axial vector meson tower. However, it is unnatural and an artifact of the large $N_c$ limit, since in reality the meson width gets broader for higher resonances and
the higher mesons participate with higher energy but also with more involved chiral interactions. So, here we just point out that axial vector meson condensation has a tendency to further reduce the total energy density, and the contribution from the tower of the resonances would not be negligible.

\section{Summary and discussion}

For searching a QCD ferromagnetism at high density of neutrons, we studied the simplest chiral Lagrangian (\ref{totalLag}) which accommodates 
neutrons at high density, 
the pion condensation, the constant magnetic field with its self energy and the QCD anomaly. 
The pion condensation is a linear spatial profile of
the neutral pion (\ref{DN}) {\it a la} Dautry and Neyman \cite{DN}
which generates a neutron spin alignment.

We solved a self-consistent equation for the total energy density for 
a given neutron density, by considering
the neutron Fermi energy, the pion self energy and also the self energy of the 
constant magnetic field.
We have shown that the minimization of energy under the assumption of spatial homogeneity leads to the ferromagnetic order preferred compared to the ordinary 
neutron
matter without the pion condensation, at the neutron density $\rho > 5 \rho_0$ where $\rho_0$ is the standard nuclear density. The result
is summarized in Fig.\ref{figano}. The generated magnetic field (see Fig.~\ref{figB}) 
is $\sqrt{eB}\sim 40$[MeV] which is around ${\cal O}(10^{17})$[G].

However, a comparison to the ALS (alternating layer structure) phase 
\cite{Tamagaki:1976ds,Takatsuka:1977uj,Takatsuka:1978ku}, which is with 
another neutral pion condensation providing a spatially alternating spin order,
shows that our ferromagnetic order has a larger energy density and thus is not favored (see Sec.~\ref{section:ALS}).

We further included axial vector mesons in our model, since the axial vector meson
condensation has the same coupling as the Dautry-Neyman neutral 
pion condensation. We found that the axial vector meson enhances the energy reduction of the ferromagnetic phase significantly (see Sec.~\ref{section:axial}).
In QCD there exists a tower of axial vector meson resonances, and inclusion of
the tower further enhances the reduction, which we roughly evaluated with the
use of the AdS/CFT correspondence (Sec.~\ref{section:axial2}).

We can summarize our results as follows:
\begin{itemize}
\item The simple chiral model with the linear neutral pion condensation and magnetic field accommodates a ferromagnetic order.
\item The QCD anomaly term lowers the ferromagnetic energy.
\item The axial vector meson condensation further reduces the energy significantly.
\item Our analysis is among spatially homogeneous phases, and needs to be compared in more detail with inhomogeneous phases such as the ALS.
\end{itemize}

Our study is based on the simple chiral model (\ref{totalLag}), so the numerical
results presented in this paper is not suitable for a detailed comparison. For example, 
inclusion of realistic nuclear forces and nucleon contact terms would give more corrections. Nevertheless, in our analysis,
in particular the axial vector meson condensation is an interesting and novel feature,
and a further consideration would be of worth. In the condensed phase of the axial vector mesons, low energy propagation modes are of interest, in view of recent progress \cite{Watanabe:2014qla} 
in non-relativistic Nambu-Goldstone theorem \cite{Watanabe:2012hr,Hidaka:2012ym}.

In this paper, we concentrated on the hadron phase,\footnote{
See for a recent attempt without a pion condensation, \cite{Diener:2010hb,Diener:2013nea,Diener:2013vza}.
} 
not a quark phase such as 
the color superconductivity. It would be interesting to
extend our calculation, if possible, to a hadron-quark mixed phase.
For a quark matter, the possibility of the ferromagnetism was studied in
\cite{Tatsumi:1999ab,Tatsumi:2011tu}, 
while the quark-hadron mixture phase was studied
in the context of neutron stars \cite{Masuda:2012kf,Masuda:2012ed}.
The high density phase of QCD still leaves a large room to be discussed
\cite{Fukushima:2010bq}, and observations of the magnetars \cite{Enoto,Ref2}
and the magnetic fields
there should reveal more about the mystery of the high density phase.


\vspace*{10mm}
{\noindent \bf Acknowledgment.} ---
The author is indebted to Tetsuo Hatsuda for the collaboration of this work at its early stage. The author would like to thank
Teruaki Enoto, Minoru Eto, Deog-Ki Hong, Kei Iida, Muneto Nitta, 
Yudai Suwa and Toru Tamagawa 
for valuable discussions and comments.
This research was
partially supported by the RIKEN iTHES project.


\end{document}